\documentclass[11pt]{article}
\usepackage{cite}
\usepackage{multicol}
\usepackage{graphicx}
\usepackage{amsmath}
\usepackage[a4paper]{geometry}
\usepackage{color,xcolor}
\definecolor{deepfuchsia}{rgb}{0.76, 0.33, 0.76}
\definecolor{pinegreen}{HTML}{008B72}

\usepackage[colorlinks=true,linkcolor=pinegreen,citecolor=magenta,urlcolor=blue]{hyperref}
\usepackage{hyperref}
\usepackage{rotating}                       
\usepackage{multirow}
\usepackage{multicol}
\usepackage{geometry}

\usepackage[para]{footmisc}

\newcommand{\la}{\langle}
\newcommand{\ra}{\rangle}

\begin{document}

\vspace*{-10mm}

\begin{center}
  {\Large \bf Kinetic freeze‑out and diffusion dynamics in small‑system \\
    asymmetric collisions at $\sqrt{s_{NN}}=200~{\rm GeV}$ in light of
    \vspace*{2mm} \\ a generalized Fokker‑Planck distribution} 
\vskip1.0cm
M.\ Waqas$^{1}${\footnote{waqas\_phy313@yahoo.com, 20220073@huat.edu.cn}},
Wolfgang Bietenholz$^{2}$ {\footnote{wolbi@nucleares.unam.mx}}, 
Khusniddin K.\ Olimov$^{3,4}${\footnote{khkolimov@gmail.com\\}},
Muhammad Ajaz$^{5}${\footnote{ajaz@awkum.edu.pk}},\\
Jihane Ben Slimane$^{6}${\footnote{jehan.saleh@nbu.edu.sa}}, 
Laila A.\ Al-Essa$^{7}${\footnote{Laalessa@pnu.edu.sa}} and
A. Haj Ismail$^{8}${\footnote{a.hajismail@ajman.ac.ae}}
\\ \ \\
   {\small\it $^1$ Hubei Key Laboratory of Energy Storage and Power Battery,
     School of Optoelectronic Engineering,\\
     School of New Energy, Hubei University of Automotive Technology,
     Shiyan 442002, China\\
     $^2$ Instituto de Ciencias Nucleares, Universidad Nacional Aut\'onoma
     de M\'exico, \\  Apartado Postal 70-543, CdMx 04510, Mexico\\
     $^3$ Physical-Technical Institute of Uzbekistan Academy of Sciences,
     100084 Tashkent, Uzbekistan,\\
     $^4$ Department of Physics, Abdul Wali Khan University Mardan,
     23200 Mardan, Pakistan\\
     $^5$ Department of Physics, Faculty of Science,
     University of Tabuk, Tabuk 47913, Saudi Arabia\\
     $^6$ Department of Physics, College of Science,
     Princess Nourah bint Abdulrahman University, \\
     P.O. Box 84428, Riyadh 11671, Saudi Arabia\\
     $^7$ Department of Mathematical Sciences, College of Science,\\
     Princess Nourah bint Abdulrahman University, P.O.Box 84428,
     Riyadh 11671, Saudi Arabia\\
     $^8$ College of Humanities and Sciences, Ajman University,
     Ajman P.O. Box 346, United Arab Emirates\\} \end{center}

\noindent

A generalized Fokker-Planck solution is used to examine the
transverse momentum ($p_{\rm T}$) spectra of neutral pions generated
in small-system asymmetric collisions, $p$-Al, $p$-Au, $d$-Au,
and $^3$He-Au, at $\sqrt{s_{NN}}=200$ GeV. This framework provides
a cohesive explanation of particle production over a broad range
of transverse momenta. We extract the energy scale governing the
transition between a thermal and a hard regime,
the effective temperature ($T$),
and the exponents determining the high-momentum falloff from fits
to PHENIX data. $T$ increases systematically with the collision
centrality and colliding system size, ranging from about 0.33 GeV
in peripheral $p$-Al collisions to 0.45 GeV in central $^3$He-Au
collisions. This increase is correlated with the average
number of participant nucleons, $\la N_{\rm part}\ra$,
and the charged‑particle pseudorapidity density,
$\la dN_{\rm ch}/d\eta \ra$, indicating
that larger and more central collisions create a denser, more
strongly interacting medium that freezes out at a higher temperature.
The acquired transition scale and power-law exponents follow consistent
patterns across systems and centralities, revealing details about
the sharpness of the transition from thermal to hard processes, and
the relative strength of momentum-space diffusion versus drag.
Interestingly, when the gold target dominates the collision geometry
in the largest system ($^3$He-Au), the transition scale becomes nearly
independent of centrality, signifying saturation of the diffusion
process. Our findings demonstrate that the generalized Fokker-Planck
solution is a sensitive probe of transport properties and non-extensive
dynamics in the quark-gluon plasma produced even in small-system
relativistic collisions, and it consistently describes pion spectra
in this set of collisions.\\

\noindent
{\bf Keywords:} momentum-space diffusion, systematic drag,
charged-hadron multiplicity, effective temperature,
generalized Fokker-Planck solution

\vskip1.0cm

\begin{multicols}{2}

\section{Introduction}
\label{sec:intro}

Quarks and gluons are usually bound to form hadrons by the strong
force. Quantum Chromodynamics (QCD) is the theory of the strong
interaction, which exhibits asymptotic freedom at short distances
and confinement at long distances \cite{asymfree,FGML,QCD50}.
The properties and composition of many hadrons have been explained
by QCD over the years \cite{Schmidt:2017bjt,Haegler,PDB,FLAG}.
According to the theory, matter is made up of six quark flavors
and eight gluons that carry the strong force between them. QCD differs
from Quantum Electrodynamics (QED)
in that gluons are non-Abelian gauge bosons, which
makes them interact with each other. This leads to
asymptotic freedom: the strong force becomes
weak over very short distances which correspond to very
high energies. Therefore, confinement breaks down when matter is
heated to a high enough temperature or compressed to a high enough
baryon density.
This leads to a deconfined state of matter (quarks, anti-quarks, and
gluons) known as Quark-Gluon Plasma (QGP). 

The QGP filled the universe just a few microseconds after the Big Bang.
At present, recreating and studying QGP in a laboratory is one of the
primary objectives in experimental high-energy physics. One can
temporarily (for about $10^{-23}~{\rm sec}$) create a hot,
dense medium that corresponds to a QGP, by colliding heavy ions
at ultra-relativistic speeds at facilities
like the Relativistic Heavy Ion Collider (RHIC)
\cite{STAR:2019bjj,STAR:2003fka} or the Large Hadron Collider (LHC)
\cite{ALICE:2019nrq,ALICE:2012ovd}.
Asymmetric collision systems, such as
$p$-Al, $p$-Au, $d$-Au, and $^3$He-Au studied at RHIC, have
also been shown to produce droplet‑like QGP signatures, offering
complementary insights into the onset of the collective behavior. These
collision experiments enable us to measure its formation, fluid-like
flow, and subsequent hadronization.

The insight about the QGP is not only due to the experimental
colliders; similar conditions may also exist
inside neutron stars. A novel approach to studying matter at
extreme temperatures and densities has emerged thanks to the first
multi-messenger observation of a binary neutron star merger,
denoted as GW170817 \cite{Abbott1,Abbott2}.

Establishing the phase diagram of strongly interacting matter is a
complicated task which incorporates experimental data, lattice
simulations \cite{Laermann:2003cv},
and QCD theory \cite{PHOBOS,PHENIX,STAR,BRAHMS}.
As an intriguing feature, the location of a Critical
Endpoint is one of the primary questions. It is believed that when
the baryon density reaches a certain level, the smooth transition
(cross-over) between conventional hadronic matter and the QGP will
turn into an abrupt, first-order phase transition. 

The QGP itself cannot be directly examined, but
its properties can be studied from a variety of experimental signatures.
Analyzing the Transverse Momentum Distributions (TMDs) of particles
created in high-energy collisions is a particularly conclusive diagnostic
technique. The shape of these momentum spectra contains relevant
information about the condition of the hot, dense
matter from which particles emerge
\cite{Adamczyk16,STAR:2015vmv,STAR:2016uxt,STAR:2017akg}. It is
possible to extract crucial thermodynamic and hydrodynamic features, such
as temperature, pressure, and energy density, by using phenomenological
models to analyze the TMDs. These parameters are essential for
constraining the Equation of State (EoS) of a QGP. As a direct result
of pressure gradients in the early-stage QGP, the anisotropy seen in
the particle emission enables the determination of properties
such as the shear viscosity and momentum-transport coefficient of
the medium \cite{Tawfik:2021xfc}. Consequently, TMDs are essential
for understanding the fluid-like behavior of the QGP. Beyond
characterizing the QGP, these spectra also serve as a probe
for new physics. 

Among the numerous observables extracted from TMDs, temperature plays
a central role. In QGP research, temperature is an essential
observable; its measurement
sheds light on the energy scale and thermodynamic
evolution of the matter resulting from high-energy collisions. The
TMD of the final-state particles is the primary data source.
At various stages during the collision evolution, the system is
described by a number of different temperature concepts. The
point of departure is the initial temperature that
reflects the highest energy density achieved right after the collision.
The yields of different particle species are fixed when the system
cools to the chemical freeze-out temperature, which is the point at
which inelastic collisions end \cite{qq,qqq}. Later, when elastic
interactions cease and particle momenta are fixed, kinetic (or
thermal) freeze-out takes place; this temperature is known as the
kinetic freeze-out temperature.

The effective temperature \cite{q1,q2,q3}
occurs at the freeze-out stage, and can be obtained through fitting
the $p_{\rm T}$-spectrum. A composite parameter represents
both the average collective transverse flow velocity and the actual kinetic
freeze-out temperature, rather than a direct thermodynamic measurement.

From a theoretical point of view, the effective temperature appears
at --- or just before --- kinetic freeze-out, but after chemical freeze-out.
The main difference is that the effective temperature comprises both
thermal motion and flow, whereas the kinetic freeze-out temperature
only represents thermal motion at decoupling. The effective temperature
can be employed to distinguish these two components as long as particles
are examined within the same centrality bins as we investigated
previously \cite{Waqas:2022fnl, Badshah:2024pyt}. A key point
is that the initial temperature is at the top of the
hierarchy, followed by the chemical freeze-out temperature and then
kinetic freeze-out temperature. In this work, a Fokker--Planck
distribution is applied to the $p_{\rm T}$-spectra of $\pi^0$-mesons
in order to extract the effective temperature and other relevant
characteristics.  

Understanding how the initial energy deposited in the collision
evolves into the final observed state of particles is an essential
objective in the study of collisions of asymmetric systems, such
as $p$–Al, $p$-Au, $d$-Au, and $^3$He-Au. The initial temperature
($T_i$) and the effective temperature ($T$) are important,
complimentary benchmarks in this exploration. The initial temperature,
which represents the peak thermodynamic condition of the developing
fireball, offers a picture of the energy density at the time of
thermalization.

On the other hand, the effective temperature derived
from the transverse momentum spectra is a composite final-state
observable that captures both the collective transverse flow
velocity that emerged throughout the system's expansion and the
actual kinetic freeze-out temperature. As a result, a great deal
of information regarding the lifetime of the system, the effectiveness
of converting initial energy into collective motion, and the total
cooling history from creation to freeze-out is encoded in the
relationship between $T_i$ and $T$.  

Section \ref{sec:method} gives a practical demonstration of the
Fokker‑Planck distribution framework. Section \ref{sec:results}
presents the resulting freeze-out parameters and the related
observables. In Section \ref{sec:conclu}
we summarize these results and discuss their implications.

\section{The method and formalism}
\label{sec:method}

The transverse momentum spectra of particles can be suitably define
by applying the Fokker-Planck framework. This technique simulates
the stochastic evolution of particles within a collision fireball,
where random scattering (diffusion) and systematic drag (drift)
interact to change the particles' momenta. The equation can be
written as \cite{Svetitsky88,Banerjee88}
\begin{eqnarray}
\label{eq1}
&& \frac{\partial P(u, t)}{\partial t} = \nonumber \\
&& \frac{\partial}{\partial u}
\Big[ A(u) P(u, t) \Big] + \frac{\partial^2}{\partial u^2}
\Big[ B(u) P(u, t) \Big], \qquad
\end{eqnarray}
where $P(u,t)$ is the probability distribution, $A(u)$ is the drift
coefficient, $B(u)$ is the diffusion coefficient, $u$ represents
the variable of interest (such as energy), and $t$ is time.

To characterize the final observed particle spectra, we search for
a stationary solution, $\partial P_{\rm s} / \partial t = 0$, which
takes the form 
 \begin{eqnarray}
\label{eq2}
P_{\rm s}(u) \propto \frac{1}{B(u)} \exp\left[-\int^{u} \frac{A(u')}{B(u')} \,
\mathrm{d}u' \right] .
\end{eqnarray}
The particular choices made for the drift coefficient $A(u)$ and the
diffusion coefficient $B(u)$ establish the
form of the stationary distribution $P_{\rm s}(u)$. The ``mixing diffusion''
scenario, in which these coefficients are presumed to depend linearly
and quadratically on the transverse kinetic energy
$E_{\rm T} = \sqrt{p_{\rm T}^{2}+m^2} - m$,
respectively, is an exceptionally insightful model
($m$ is the particle mass).
This scenario assumes
\begin{eqnarray}
\label{eq3}
A(E_{\rm T}) &=& A_0 + \alpha E_{\rm T} \ , \nonumber\\
B(E_{\rm T}) &=& B_0 + \beta E_{\rm T}^2 \ .
\end{eqnarray} 
Substituting these forms into the general stationary solution yields
a closed-form expression
\begin{eqnarray}
\label{eq4}
P_{\rm s}(E_{\rm T}) = C \, \frac{\exp\left[-\displaystyle\frac{b}{T}
\arctan\left(\frac{E_{\rm T}}{b}\right)\right]}{\left[1 +
\left(\displaystyle\frac{E_{\rm T}}{b}\right)^{2}\right]^{c}} \ .
\end{eqnarray}
The additional phenomenological parameters $b$, $T$, and $c$ in
eq.\ (\ref{eq4})
are related to the model's fundamental coefficients by
\begin{equation} \label{parameq}
b = \sqrt{B_0 / \beta}\ , \ \ T = B_0/A_0 \ , \ \
c = 1 + \alpha / (2\beta) \ ,
\end{equation}
while $C$ acts as an overall normalization constant.
The function $P_{\rm s}(E_{\rm T})$  is a suitable
contender to characterize the entire transverse momentum spectrum
because it smoothly joins a low-$p_{\rm T}$ exponential regime to
a high-$p_{\rm T}$ power-law tail. 

Equation (\ref{eq4}) is appropriate for characterizing particle spectra
due to its mathematical behavior at its limits. The function
reduces to a power-law, $P_{\rm s}(E_{\rm T}) \propto p_{\rm T}^{-2c}$, in the
high transverse momentum domain, where $p_T \gg b$
or equivalently $E_{\rm T}/b \gg 1$.
On the other hand, it reduces to a Boltzmann-like
exponential at low momentum ($p_{\rm T} \ll b$ or $E_{\rm T}/b \ll 1$),
$P_{\rm s}(E_{\rm T}) \propto e^{-E_{\rm T}/T}$. Both the Tsallis distribution
and the particle spectra commonly seen in heavy-ion collisions
exhibit these two asymptotic forms: an exponential shape at low
$p_{\rm T}$ and a power-law tail at high $p_{\rm T}$.

The Fokker-Planck solution was initially used to investigate particle
spectra in central Pb-Pb collisions at $\sqrt{s_{NN}}= 2.76~{\rm TeV}$
\cite{HZheng}. However, the denominator's power was 4 instead of 2
in that original study. Further extensive investigation
\cite{XuejiaoYin} showed that a more flexible form was needed
to properly describe the spectra of observed particles across all
centralities. As a result, a generalized Fokker-Planck solution
was adopted \cite{ZZZB20},
\begin{equation}
\label{eq5}
\left. E_{\rm T} \frac{\mathrm{d}^{3}N}{\mathrm{d}p^{3}} \right|_{|\eta| < a}
= C \, \frac{\exp\left[-\displaystyle\frac{b}{T} \arctan
\left(\frac{E_{\rm T}}{b}\right)\right]}
{\left[1 + \left(\displaystyle\frac{E_{\rm T}}{b}\right)^{d}\right]^{c}} \ .
\end{equation}
The parameter $a=0.35$ determines the measurement's pseudorapidity
acceptance. In accordance with the experiment's pseudorapidity coverage,
it defines the interval $|\eta|$ $<$ $a$ within which particles are
observed. The required flexibility is provided by the five parameters
$C$, $b$, $c$, $d$, and $T$. The characteristic energy $E_{\rm T}$,
at which the spectrum switches from an exponential to
a power-law behavior, is specified by the parameter $b$.
The effective temperature scale is symbolized by $T$.
The high-momentum tail of the spectrum is influenced by the
combination of the exponents $c$ and $d$.
At high energy, the spectrum follows a power-law $\propto E_{\rm T}^{-cd}$.
When combined, these features enable the distribution
to describe both the hard-scattering (high $p_{\rm T}$) as well as the
thermal (low $p_{\rm T}$) regime of particle generation.

While $b$, $c$, and $d$ contain information about the dynamics of
the collision medium, such as the strength of diffusion and
non-extensive effects originating from the Fokker-Planck transport
process, $T$ is typically considered as the freeze-out temperature
in a freeze-out analysis. It should be emphasized that the parameters
$C$, $b$, $c$, $d$, and $T$ are not directly measurable;
they are fitting parameters whose physical interpretation is obtained
in the Fokker‑Planck framework. The structure of the denominator,
specifically the term raised to the exponent $c$, is responsible for
the power-law tail seen at higher energies. This behavior, in which
the spectrum declines as a power-law $\propto E_{\rm T}^{-cd}$, means a
deviation from conventional Boltzmann statistics.
Such a form is frequently linked to non-extensive
thermodynamics, describing systems that exhibit memory effects or
long-range interactions, the influence of which is captured by the
exponents $c$ and $d$. In a similar manner, the diffusion strength
that arises in the original Fokker-Planck transport process is
integrated into the reference energy $b$.

The energy scale at which the spectrum
transforms from an exponential-like form to a power-law is defined
in part by the diffusion coefficient throughout the derivation.
Consequently, the fitted value of $b$ effectively reveals the
transition energy, and provides information on the medium's
stochastic momentum scattering intensity.

In addition to $b$, the
exponents $c$ and $d$ establish a phenomenological link to the
collision dynamics. The thermal slope of the spectrum at
low transverse momentum is described by $T$, which is the
kinetic freeze-out temperature.

\section{Results and discussion}

\label{sec:results}

\begin{figure*}
\centering
 \includegraphics[width=0.6\textwidth]{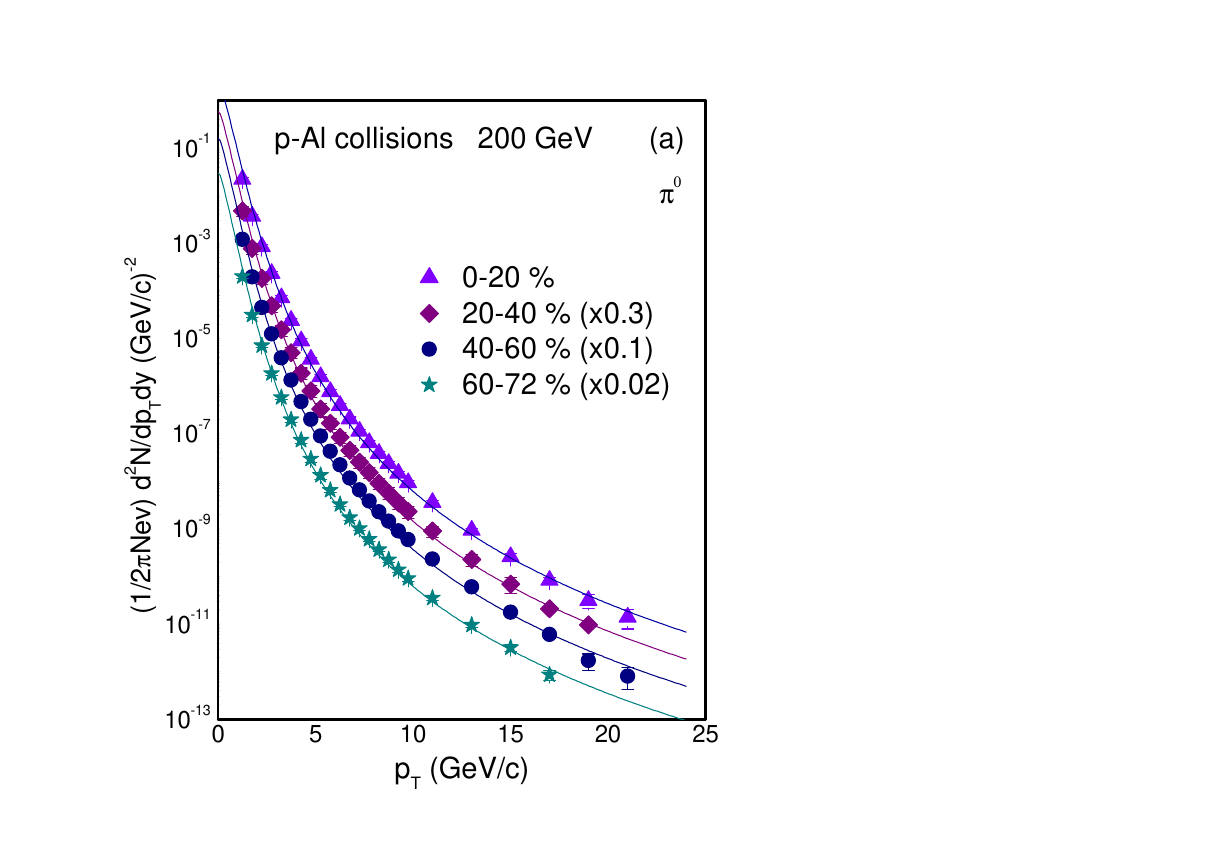} \hspace{-4cm}
 \includegraphics[width=0.6\textwidth]{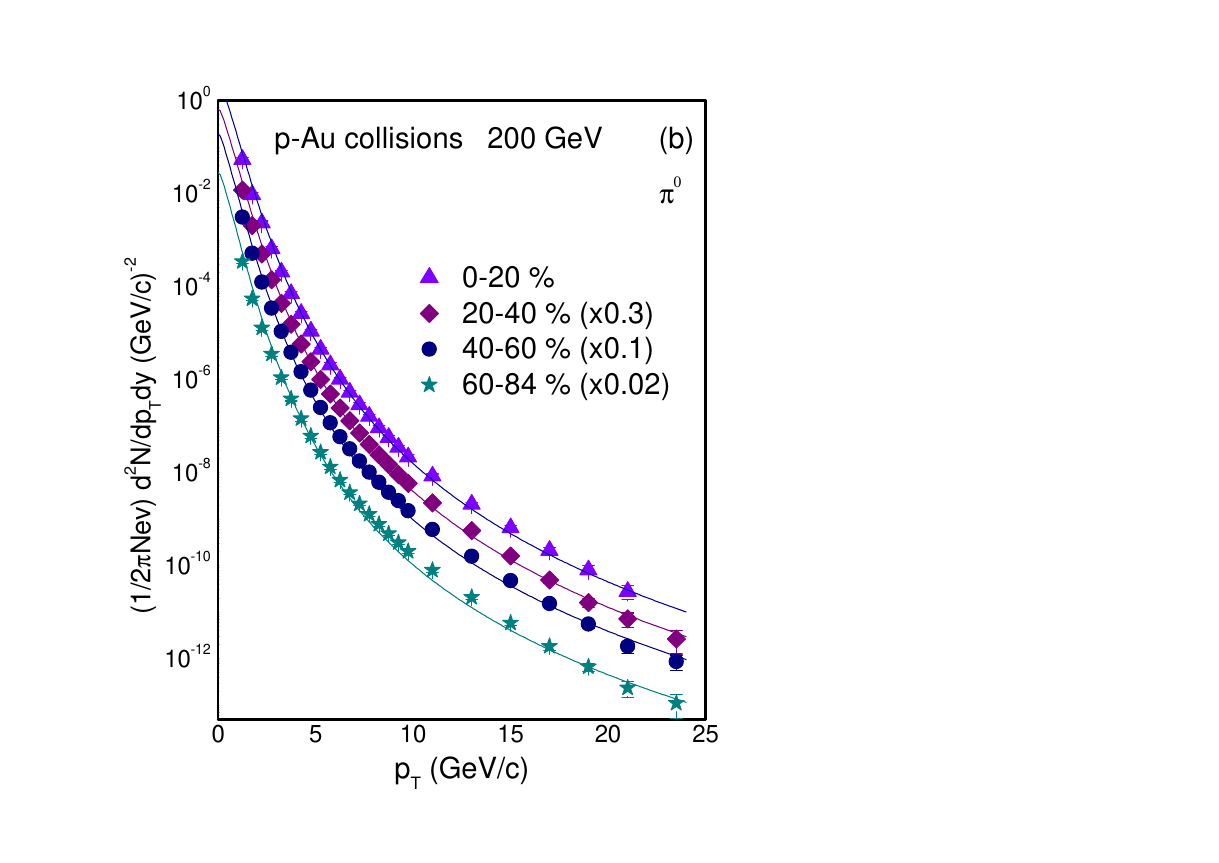} \\
 \includegraphics[width=0.6\textwidth]{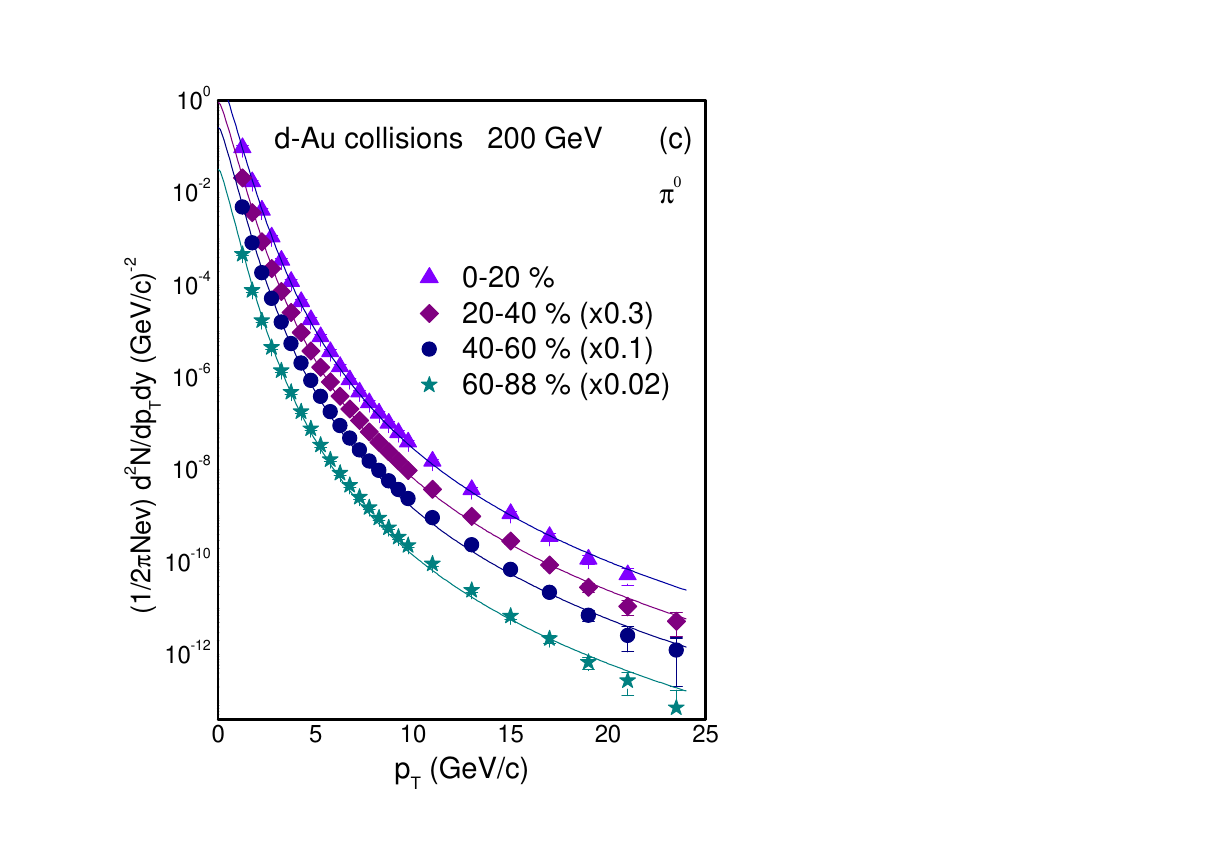} \hspace{-4cm}
 \includegraphics[width=0.6\textwidth]{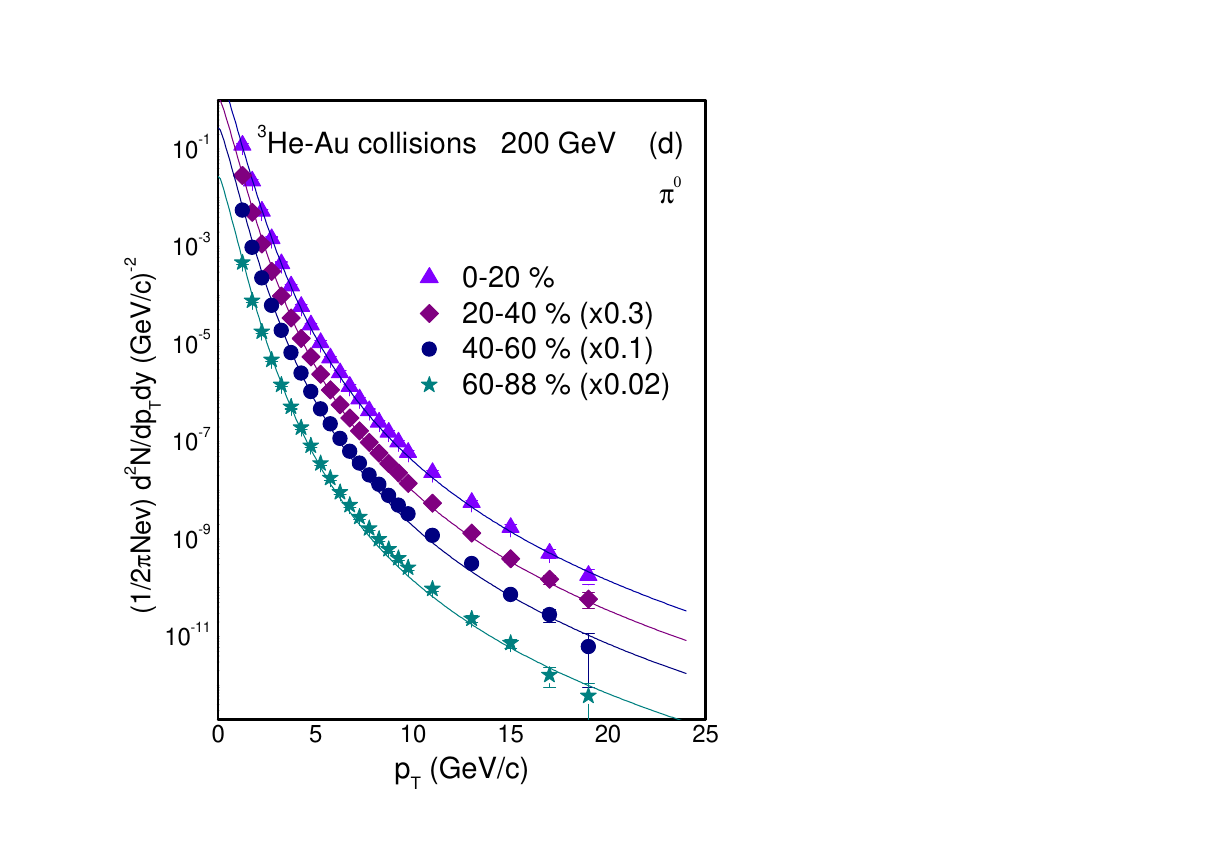}
\caption{The $p_{\rm T}$-spectra of neutral pions measured in $p$-Al,
$p$-Au, $d$-Au, and $^3$He-Au collisions at $\sqrt{s_{NN}}=200~{\rm GeV}$
are shown together with fits based on the generalized Fokker-Planck
distribution (\ref{eq5}).
$N_{\rm ev}$ is the number of events, $N$ the number of $\pi^{0}$-mesons,
and $y$ is the rapidity.
The data are taken from the PHENIX Collaboration \cite{PHENIX:2021dod}.} 
\label{fig1}
\end{figure*}

Figure \ref{fig1} displays the transverse momentum ($p_{\rm T}$)
distributions of neutral pions ($\pi^0$), measured in four distinct
collision systems at the same center-of-mass energy of
$\sqrt{s_{NN}} = 200~{\rm GeV}$.
Each panel corresponds to one system: (a) proton-Aluminium
($p$-Al), (b) proton-Gold ($p$-Au), (c) deuteron-Gold ($d$-Au),
and (d) Helium-Gold ($^3$He-Au) collisions. The generalized Fokker-Planck
distribution (\ref{eq5}) \cite{Svetitsky88,ZZZB20,NJP,Yin:2017pzp}
is used to fit the published $\pi^0$ transverse momentum spectra
from these four collision systems \cite{PHENIX:2021dod}.
The fitted Fokker-Planck curves reveal a consistent behavior
across all four systems. A key outcome is that the same functional
form successfully matches the data over the entire transverse
momentum range, in each centrality class; no changes to the
model's core expression are required. This indicates that the
underlying transport mechanism described by the Fokker-Planck
equation remains applicable regardless of collision size or
centrality.

The curves also exhibit a clear centrality dependency:
spectra from central collisions are systematically harder and
less steep at low $p_{\rm T}$, while peripheral collisions produce
steeper, more sharply falling distributions. This smooth transition
is fully described by adjusting only the model's parameters for
each centrality bin. Moreover, a steady hardening of the spectrum
is evident when moving from lighter to heavier collision systems.
Within the same centrality class, the fitted curve becomes progressively
flatter from $p$-Al to $^3$He-Au. Again, this trend is captured
entirely through parameter tuning, without altering the fundamental
Fokker-Planck form. The parameters extracted from our model along
with the values of the ratio $\chi^2$/dof are tabulated in Table \ref{tab1}. 

\begin{figure*}
\centering
  \includegraphics[width=1.\textwidth]{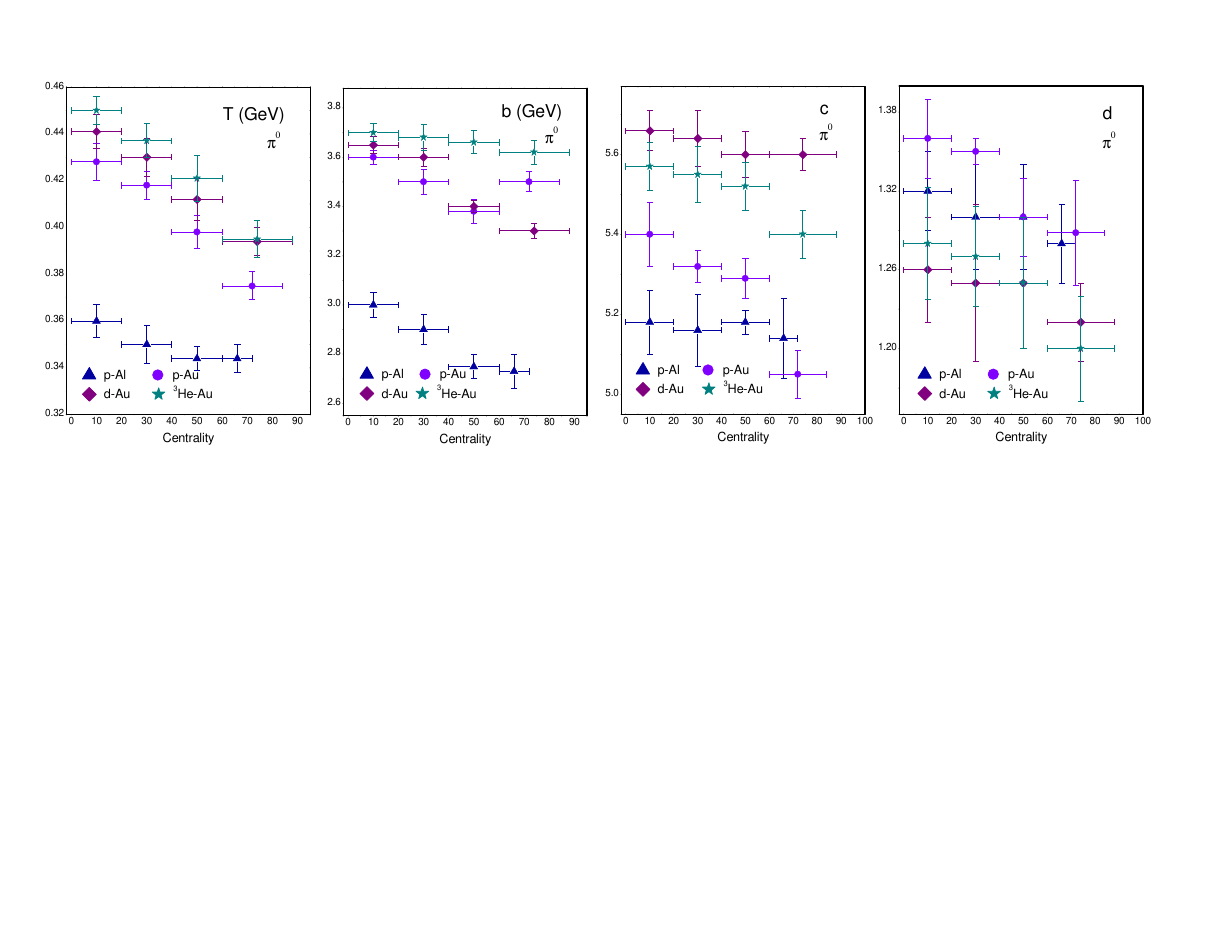}
\vspace*{-7cm}  
\caption{The four panels illustrate the fitting results of the
parameters $T$, $b$, $c$, and $d$, each one in four centrality
intervals of the collisions $p$‑Al, $p$‑Au, $d$‑Au and $^3$He-Au.} 
\label{fig2}
\end{figure*}

\begin{table*}
{\small
\begin{center}
\begin{tabular}{cc|ccccccccc}\\ \hline
  Collision & Centrality  & $C$  & $b~~({\rm GeV})$  & $c$ & $d$
  & $T~({\rm GeV})$ & $\chi^2$/dof \\
\hline
\multirow{4}{*}{$p$-Al} & $0-20\%$ & $9.84\pm0.7$ &$3\pm0.05$
& $5.18\pm0.08$ & $1.32\pm0.03$    & $0.360\pm0.007$   & 142/19\\
& $20-40\%$    & $8.05\pm0.66$     & $2.9\pm0.06$
& $5.16\pm0.09$  & $1.30\pm0.04$   & $0.350\pm0.008$   & 149/18\\
& $40-60\%$    & $6.468\pm0.75$    & $2.75\pm0.05$  & $5.18\pm0.03$
& $1.30\pm0.04$      & $0.344\pm0.005$   & 149/19\\
& $60-72\%$    & $3.337\pm0.36$    & $2.73\pm0.07$  & $5.14\pm0.1$
& $1.28\pm0.03$      & $0.334\pm0.006$   & 112/17\\
\hline
\multirow{4}{*}{$p$-Au} & $0-20\%$ & $12.2\pm1$ & $3.6\pm0.03$  &
$5.4\pm0.08$ & $1.36\pm0.03$    & $0.428\pm0.008$  & 114/19\\
& $20-40\%$    & $10.28\pm0.8$ &$3.5\pm0.05$  & $5.32\pm0.04$
& $1.35\pm0.01$ & $0.418\pm0.006$      & 163/20\\
  & $40-60\%$    & $8.5\pm0.77$ &$3.38\pm0.05$  & $5.29\pm0.05$
& $1.3\pm0.03$ & $0.395\pm0.007$      & 111/20\\
  & $60-84\%$    & $6.185\pm0.8$ &$3.5\pm0.04$  & $5.05\pm0.06$
& $1.28\pm0.04$ & $0.375\pm0.006$      & 171/20\\
\hline
\multirow{4}{*}{$d$-Au} & $0-20\%$ & $16.2\pm1.14$ &$3.65\pm0.033$
& $5.6\pm0.05$  & $1.26\pm0.04$    & $0.441\pm0.0074$  & 42/20\\
& $20-40\%$    & $14.04\pm1.17$ &$3.6\pm0.037$  &
$5.64\pm0.07$ & $1.25\pm0.06$ & $0.430\pm0.008$      & 53/20\\
& $40-60\%$    & $11.73\pm0.9$ &$3.4\pm0.025$  &
$5.6\pm0.058$  & $1.25\pm0.05$ & $0.412\pm0.009$      & 88/20\\
& $60-88\%$    & $6.75\pm0.88$ &$3.3\pm0.03$  &
$5.6\pm0.04$ & $1.22\pm0.03$ & $0.394\pm0.006$      & 107/20\\
\hline
\multirow{4}{*}{$^3$He-Au} & $0-20\%$ & $18.06\pm2.4$ &$3.7\pm0.035$  &
$5.57\pm0.06$  & $1.28\pm0.043$    & $0.450\pm0.006$      & 51.5/18\\
 & $20-40\%$    & $17.43\pm2.3$ &$3.68\pm0.052$  &
$5.55\pm0.07$  & $1.27\pm0.038$    & $0.437\pm0.0072$ & 59/18\\
 & $40-60\%$    & $12.6\pm1.27$ &$3.66\pm0.047$  &
$5.52\pm0.06$  & $1.25\pm0.05$    & $0.421\pm0.010$      & 121/18\\
 & $60-88\%$    & $6.068\pm0.84$ &$3.62\pm0.05$  & $5.4\pm0.06$
& $1.2\pm0.04$    & $0.395\pm0.008$  & 118/18\\
\hline\\
\end{tabular}
\end{center}}
\vspace{-3mm}
\caption{Values of the parameters $C$, $b$, $c$, $d$, and $T$
($C$ is the normalization constant that adjusts the fit curve
to the experimental data), and $\chi^2$ per degree of freedom (dof),
corresponding to the curves in Figure \ref{fig1}.\label{tab1}}
\vspace{-6mm}
\end{table*}

Figure \ref{fig2} displays the values obtained for the parameters
depending on centrality, as well as their comparisons in the four
asymmetric systems. The four panels
illustrate the result for the effective temperature $T$,
as well as the parameters $b$, $c$, and $d$, respectively.
The trend of these symbols, from left to right,
shows their centrality dependency. 

The first panel shows systematic patterns that provide
insight into the thermal nature of the generated fireball
by analyzing the effective temperature $T$ obtained from
the Fokker-Planck fits. $T$ progressively increases
as the collisions become more central in each of these four collision
systems. The extracted values of $T$ are least 
and attain $0.334 \pm 0.006$ for peripheral ($p$-Al) collisions,
where the projectile and target have minimal geometric overlap.
In the most central events, the temperature increases consistently to
$T \approx 0.450 \pm0.006~{\rm GeV}$ ($^3$He-Au) as centrality increases,
indicating a larger overlap region and greater deposited energy.
This rise demonstrates how, at the moment of kinetic freeze-out,
a more extended and actively interacting medium creates a hotter
thermal state. The results show a hierarchy among the collision systems,
in addition to the clear centrality dependency:
$p$-Al collisions have the lowest $T$ at a given centrality. The highest
$T$ values are consistently found in $^3$He-Au, followed by $d$-Au,
then $p$-Au. When moving from a single proton to a deuteron and
subsequently to a helium-3 projectile, as well as from an aluminium
to a gold target, this ordering corresponds to the increase in the
number of participating nucleons. More participants generally result
in stronger radial flow, more frequent re-scattering, and higher
initial energy density, all of which increase the temperature $T$ at
the kinetic freeze-out stage. This parameter explicitly reflects the
changing thermal environment governed by the collision geometry, as
demonstrated by the smooth, monotonic rise of $T$ with centrality,
which is consistent for the different systems.
Notably, the ratio $B_0/A_0$ of the baseline diffusion and drift
coefficients determines $T$ in the Fokker-Planck formulation,
see eq.~(\ref{parameq}). Therefore, when the
medium becomes denser and more interacting, the apparent rise of $T$
with centrality can be understood as a relative strengthening of
momentum-space diffusion compared to drag.

Our previously published measurements \cite{Waqas:2022fnl} of
effective temperatures for $K^+$-mesons
at RHIC energies are consistent with the numerical values obtained
here (covering approximately 0.334 to 0.450 GeV), which strengthens
the physical relevance of the model.  The trend of $T$ also agrees
with Refs.\ \cite{Waqas:2022nae,Lao:2021wub,Waqas:2021rmb}. Overall,
the effective temperature not only reflects the system's anticipated
heating in larger and more central collisions, but it also supports
the Fokker-Planck transport framework as a trustworthy method for
describing the fireball's thermal state at freeze-out.  

The behavior of the parameter $b$ with respect to centrality and its
fluctuation with system size are displayed in the second panel.
In the generalized Fokker-Planck solution, the parameter $b$ describes
the characteristic energy scale where the spectrum transitions
from exponential to power-law behavior.
With centrality and system size, this parameter
exhibits clear patterns. In the smallest system, $p$-Al, $b$
drops from $3.0\pm0.05~{\rm GeV}$ in the $0-20\%$ bin to
$2.73\pm0.07~{\rm GeV}$ in the $60 - 72\%$ bin, indicating that
the transition to a power-law tail switches to lower
transverse momenta when collisions become more peripheral.
Both $p$-Au and $d$-Au exhibit a similar,
although less pronounced decrease: for p-Au, $b$ drops from
$3.6 \pm 0.03~{\rm GeV}$ to $3.38 \pm 0.04~{\rm GeV}$
(with a slight increase in the most peripheral bin),
while for $d$-Au, it falls from $3.65 \pm 0.033~{\rm GeV}$ to
$3.3 \pm 0.03~{\rm GeV}$. On the other hand, $^3$He-Au displays nearly
constant $b$-values throughout the centralities, varying only
from $3.7 \pm 0.035~{\rm GeV}$ to $3.62 \pm 0.05~{\rm GeV}$.
This suggests that even peripheral interactions
in this largest system retain a diffusion strength that maintains
the transition energy scale essentially constant.

At a certain centrality, $b$ grows as the colliding
system becomes larger. It increases for the $0-20\%$ bin,
for instance, from $3.0~{\rm GeV}$ ($p$-Al) to $3.6~{\rm GeV}$
($p$-Au), $3.65~{\rm GeV}$ ($d$-Au), and $3.7~{\rm GeV}$ ($^3$He-Au).
As participant multiplicity and energy density rise, momentum-space
diffusion becomes stronger (as reflected by the
increase in $b$), shifting the power-law crossover
to higher $p_{\rm T}$.

The observed increase of the parameter $b$ with the
collision centrality and system size can be explained by an enhanced
transverse flow velocity and the associated pressure gradient in
the expanding system. More central collisions lead to a larger
number of participant nucleons, creating a region of higher energy
density and pressure. The enhanced pressure gradient leads to
a stronger collective expansion, known as radial flow, in more central
(as opposed to peripheral) collisions. An increased radial flow boosts
particles to higher transverse momenta, which ``flattens'' the spectrum,
shifting the spectral transition between an exponential and a power-law
behavior to higher transverse momenta. Therefore, a larger $b$-value
is required to account for this flatter slope in the low-to-intermediate
region for more central collisions.

The behavior of $b$ complements the temperature
results: $b$ tracks the location of the transition region, which
also changes to higher energies in larger and more central
collisions, whereas $T$ exhibits the thermal slope at low $p_{\rm T}$ and
increases with centrality and system size. Even when the projectile
overlap is reduced, the near-constant $b$ in $^3$He-Au across
centralities could suggest that the gold target dominates
the diffusion scale. The observation that the Fokker-Planck
transport framework consistently describes the medium's evolution
from small to large systems is strengthened by these trends taken
together.

\begin{figure*}
\centering
  \includegraphics[width=1.\textwidth]{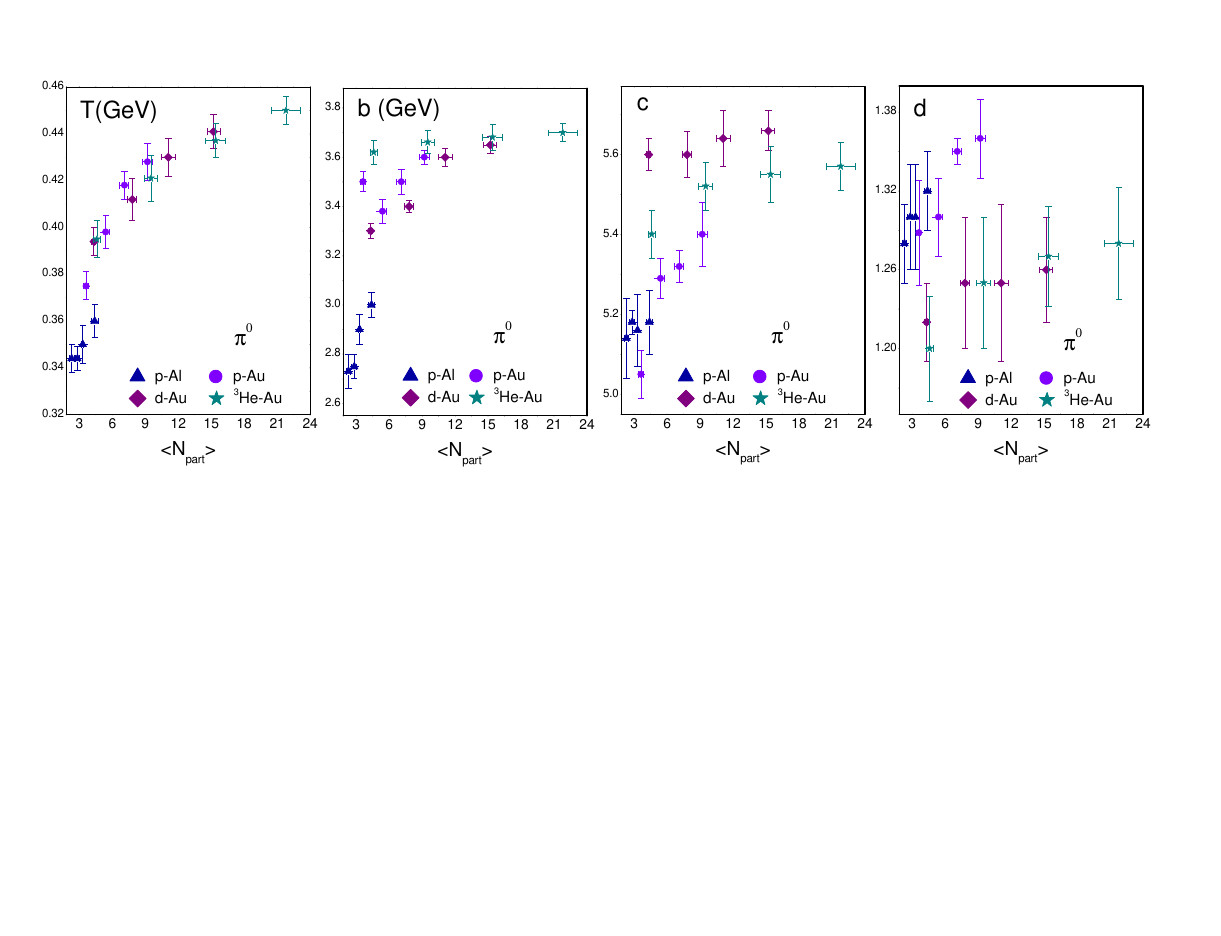}
\vspace*{-7cm}
\caption{The correlation of the average number of participant
nucleons, $\la N_{\text{part}}\ra$, with the effective
temperature $T$ and the parameters $b$, $c$, and $d$.
The data of $\la N_{\text{part}}\ra$ is taken from Ref.\ \cite{PHENIX:2021dod}.} 
\label{fig3}
\end{figure*}

The parameter $c$ exhibits a clear ordering with the system size in the
third panel. It rises from $\approx 5.18$ in $p$-Al to $5.4$ in $p$-Au
at the most central bin ($0-20\%$), then to $5.6$ in $d$-Au
and $5.57$ in $^3$He-Au. This evolution shows that at
high transverse momentum, larger projectiles coupled with a
heavy gold target produce a steeper power-law tail (since
\(P_{\rm s}(E_{\rm T}) \propto E_{\rm T}^{-c d}\)), suggesting that the
spectrum falls off more rapidly. The centrality dependency of $c$ is
system-dependent: for $p$-Al, $c$ only varies
within the error bars and is nearly constant over all
centrality bins. $p$-Au collisions yield in a more
centrality-dependent spectrum, as confirmed by the monotonous reduction
from $5.4$ ($0-20\%$) to $5.05$ ($60-84\%$). With just a slight
drop in the most peripheral bin, $c$ in $d$-Au and $^3$He-Au
remains almost constant across most centralities, indicating
that non-extensive effects are less sensitive to centrality
once the projectile is composite.

In the fourth panel, a distinct system-size hierarchy can be observed
by the parameter $d$. At $0 - 20\%$ centrality, $d$ is largest for
$p$-Au ($d=1.36$), followed by $p$-Al ($d=1.32$), $^3$He-Au ($d=1.28$),
and $d$-Au ($d=1.26$). The fact that this ordering deviates from the
monotonic increase observed in $c$ suggests that $d$ captures
a unique feature of the spectral shape, specifically the abrupt
change from the exponential low-$p_{\rm T}$ region to the power-law
tail. The centrality dependency of the parameter $d$ also varies:
in $p$-Al, it decreases only little from $1.32$ to $1.28$;
in $p$-Au, it decreases more significantly from $1.36$ to $1.28$;
in $d$-Au and $^3$He-Au, the decline is gradual, with $d$ moving
from around $1.26$ to $1.22$ and from $1.28$ to $1.20$, respectively.
Interestingly, for all centralities, the parameter $d$ in $d$-Au
collision is consistently lower than in $p$-Au. This suggests
that the presence of a bound two-nucleon projectile softens
the momentum-space diffusion process compared to a single proton,
leading to a more gradual transition between the thermal
and hard-scattering regimes.  

The first panel of Figure~\ref{fig3} shows the correlation between
the effective temperature $T$, and the average number of participant
nucleons, $\la N_{\rm part}\ra$. The data for
$\la N_{\rm part} \ra$ is taken from Ref.~\cite{PHENIX:2021dod}.
All four asymmetric collision systems display a monotonous increase
of $T$ with $\la N_{\rm part}\ra$, offering a direct
quantitative link between the collision geometry and the thermal
state of the produced medium. For all systems, $T$ increases
monotonically with $\la N_{\text{part}}\ra$.
For example, in the most
central bin, $T$ rises from 0.360~GeV in $p$-Al to 0.450~GeV
in $^3$He-Au while $\langle N_{\text{part}}\rangle$ increases
from $4.4$ to $21.8$.
We repeat that the ratio $B_0/A_0$ of the drift and diffusion
coefficients at baseline defines $T$, cf.\ eq.\ (\ref{parameq}).
Thus, the positive correlation
shows that the relative strength of momentum-space diffusion over
drag increases as the system becomes larger and more central
({\it i.e.}, as $\la N_{\rm part} \ra$ increases).
In a denser, more strongly interacting medium, random scattering
becomes more powerful compared to the systematic drag, which
is consistent with this behavior. As explained for Figure~\ref{fig2},
this correlation reflects the higher energy density and stronger
collective expansion in larger collisions.
Taken together, the patterns suggest
that the Fokker-Planck transport description is appropriate
for a broad range of collision sizes and centralities.

Similar to the effective temperature $T$, there is a positive
correlation between $\la N_{\rm part}\ra$ and the
parameter $b$ in the second panel of Figure~\ref{fig3}.
However, these two quantities capture
different features of the transport dynamics.
In the Fokker-Planck framework,
$b = \sqrt{B_0/\beta}$ isolates the ratio of the constant
part $B_0$ to the quadratic coefficient $\beta$ that appears
in the diffusion term of eq.\ (\ref{eq3}),
$B(E_{\rm T}) = B_0 + \beta E_{\rm T}^2$,
while $T = B_0/A_0$ quantifies the balance between the drift
coefficient $A_0$ and the baseline diffusion coefficient $B_0$.
Therefore, in larger and more central collisions, $B_0$ grows
relatively to both $A_0$ and $\beta$, because both $T$ and $b$
increase simultaneously with the participant number. This is
compatible with a denser, more strongly interacting medium:
$B_0$ becomes more dominant relative to $A_0$, resulting in
a higher $T$, and the constant part of the diffusion coefficient
grows faster than the quadratic part. As shown in Table~\ref{tab1}
and Figure~\ref{fig2}, $b$ generally decreases with
peripherality in smaller systems ($p$-Al, $p$-Au, $d$-Au).
A notable difference emerges in the $^3$He-Au system:
$b$ remains nearly constant across all centralities,
whereas $T$ continues to rise with
$\la N_{\rm part}\ra$. This discrepancy demonstrates that,
when the target (gold) dominates the collision geometry, the
ratio $B_0/\beta$ saturates and the constant and quadratic
components of diffusion scale simultaneously, while the ratio
$B_0/A_0$ keeps rising, suggesting an additional increase in
the relative strength of diffusion over drag. In contrast, $b$
fluctuates more strongly with centrality in the smaller systems,
indicating that the diffusion scale has not yet reached saturation.
Therefore, beyond the pure effect of the thermal temperature,
the distinct behaviors of $b$ and $T$ offer further insight into
the manner how the random component of momentum evolution fluctuates
with system size and centrality.
 
The correlation between the parameter $c$ and
$\langle N_{\text{part}}\rangle$ is illustrated in the third panel
of Figure~\ref{fig3}. Together with $d$, $c$ governs the
power-law falloff of the transverse momentum spectrum at high
$p_{\rm T}$ (more precisely, $P_{\rm s}(E_{\rm T}) \propto E_{\rm T}^{-c d}$),
cf.\ Section \ref{sec:method}.
It is linked to the drift and diffusion coefficients via
$c = 1 + \alpha/(2\beta)$, see eq.\ (\ref{parameq}). A distinct
ordering with projectile size is seen in the extracted values.
This progression shows a monotonous change
in the transport coefficients: the ratio $\alpha/\beta$ increases
with the system size, revealing that the drift term becomes more
significant in relation to the quadratic diffusion component.
Additionally, the centrality dependency of $c$ varies across
systems, as already detailed in the discussion of Figure~\ref{fig2}
and Table~\ref{tab1}. Briefly, $c$ remains nearly constant in $p$-Al
and $d$-Au across centralities, decreases systematically with
peripherality in $p$-Au, and only drops in the most peripheral
bin of $^3$He-Au. These trends indicate that the ratio
$\alpha/\beta$ (drift relative to quadratic diffusion) saturates
in small or composite projectiles, while it becomes sensitive to the
participant number only when the projectile contribution diminishes
significantly.  
\begin{figure*}
\centering
  \includegraphics[width=1.\textwidth]{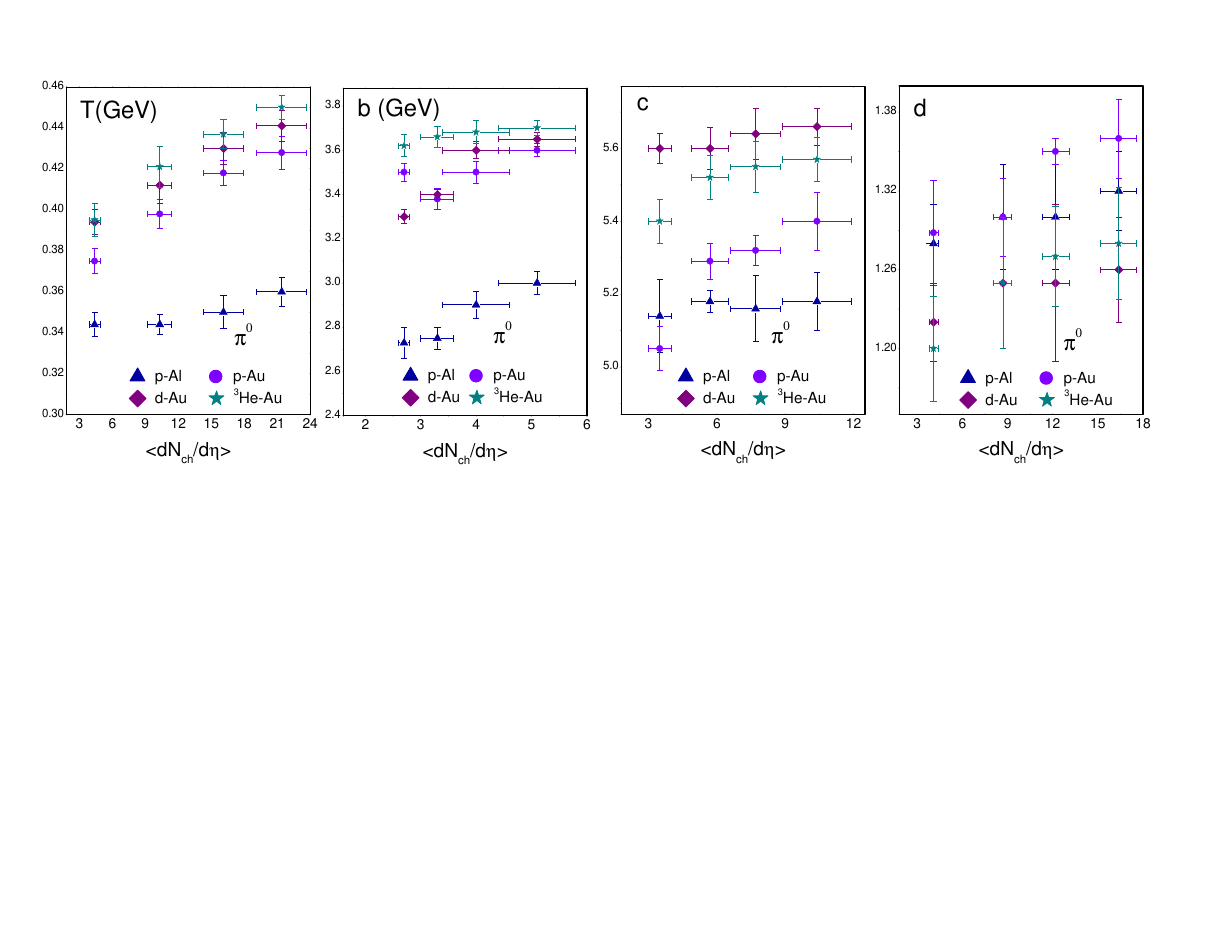}
\vspace*{-7cm}
\caption{The correlation of the charged particle multiplicity
per unit of pseudorapidity, $\la dN_{\rm ch}/d\eta \ra$, with the
effective temperature and the parameters $b$, $c$, and $d$.
The data of $\la dN_{ch}/d\eta \ra$ is taken from
Ref.\ \cite{PHENIX:2021dod}.} 
\label{fig4}
\end{figure*}

When comparing the systems at a fixed centrality, $c$ grows
monotonically with the projectile size, which is identical to the
behavior of $T$ and $b$.
This consistently supports the property that a denser medium with
modified transport coefficients is generated by larger, more
central collisions. The
distinct behaviors in the peripheral bins, a steep decline
for $p$-Au, a steady decline for $^3$He-Au, and constant values
for $p$-Al and $d$-Au, highlight how projectile size and target
dominance interact to shape the non-extensive nature of the
produced matter. Overall, by linking the power-law tail to the
ratio of drift and diffusion coefficients, the parameter $c$
provides complementary information about the underlying transport
dynamics.

Conversely, the fourth panel of Figure~\ref{fig3} illustrates
the correlation between the parameter $d$ and
$\la N_{\text{part}}\ra$. The power-law falloff of the
transverse momentum spectrum at high $p_{\rm T}$,
$P_{\rm s}(E_{\rm T}) \propto E_{\rm T}^{-c d}$, is defined by the
product $c d$, as stated before, while the
sharpness of the transition from the exponential low-$p_{\rm T}$ region
to the power-law tail is controlled by $d$. A distinct ordering
with system size is seen in the extracted values (see Table~\ref{tab1}
and the discussion of Figure~\ref{fig2}). This ordering does not follow a
monotonic increase with projectile mass: $p$-Au exhibits the
largest value of $d$, while $d$-Au and $^3$He-Au yield smaller values,
indicating that $d$ captures a distinct aspect of the spectral
shape that is not simply scaled by the participant number.

From central to peripheral collisions, $d$ decreases within
each system, as detailed in Table~\ref{tab1} and the discussion
of Figure~\ref{fig2}. This consistent reduction indicates that
the transition from exponential to power‑law behavior becomes
more gradual in peripheral collisions, where the lower density
and weaker collective expansion broaden the transition region.
When comparing systems at fixed centrality, $d$ is largest
for $p$-Au, intermediate for $p$-Al and $^3$He-Au, and smallest
for $d$-Au, revealing that $d$ is sensitive to the specific
combination of projectile and target rather than just the participant
number. The near constance of $d$ in $p$-Al across centralities
suggests saturation of the transition sharpness in this small
system, while the pronounced drop in $^3$He-Au reflects the gold
target's diminishing role in peripheral collisions. Together with
$c$, the parameter $d$ provides a complete description of the
spectral shape: the overall power‑law exponent is $c d$,
with $d$ controlling the sharpness of the transition and $c$
governing the asymptotic falloff via the ratio $\alpha/\beta$.
These patterns support the Fokker-Planck description as a unified
framework for particle production across different collision
systems and centralities. 

To further explore the system-size dependency, we compare the
extracted parameters \(T\), \(b\), \(c\), and \(d\) with the
charged-hadron multiplicity per unit of pseudorapidity,
\(\langle dN_{\rm ch}/d\eta\rangle\), in each collision system.
The data for \(\langle dN_{\rm ch}/d\eta\rangle\) is taken
from Ref.\ \cite{PHENIX:2021dod} and the measured
values are given in Table \ref{tab2}.
\begin{table*}[ht]
\begin{center}
\begin{tabular}{|c||r|r|r|r|}
\hline
\multirow{2}{*}{collision} & \multicolumn{4}{c|}{centrality} \\
\cline{2-5} 
 & 0 -- 20\% & 20 -- 40\% & 40 -- 60\% & $> 60\%$ \\
 \hline
$p$-Al & $5.1 \pm 0.7$ & $4.0 \pm 0.6$ & $3.3 \pm 0.3$ & $2.7 \pm 0.1$ \\
$p$-Au & $10.4 \pm 1.5$ & $7.7 \pm 1.1$ & $5.7 \pm 0.8$ & $3.5 \pm 0.5$ \\
$d$-Al & $16.4 \pm 1.2$ & $12.2 \pm 0.9$ & $8.7 \pm 0.6$ & $4.1 \pm 0.3$ \\
$^{3}$He-Au & $21.4 \pm 2.3$ & $16.1 \pm 1.8$ & $10.3 \pm 1.1$
& $4.4 \pm 0.5$\\
\hline
\end{tabular}
\end{center}
\caption{The values of the charged-hadron multiplicity per unit of
pseudorapidity, $\la dN_{\rm ch}/d\eta\ra$, for the four collisions under
consideration, in the same centrality intervals as in Table \ref{tab1}.}
\label{tab2}
\end{table*}

A mild but significant
positive correlation exists within each system for all parameters,
as Figure~\ref{fig4} illustrates: greater
\(\langle dN_{\rm ch}/d\eta\rangle\) corresponds to higher
$T$, higher $b$, higher $c$, and higher $d$. Likewise, each
parameter $T$, $b$ and $c$ evolves monotonically with projectile
size when comparing the four systems at a fixed centrality,
according to the ordering from
\(p\)-Al to \(p\)-Au to \(d\)-Au to \(^3\)He-Au
(the parameter $d$ shows a distinct non-monotonic behavior,
as discussed above). These patterns agree with
the previously provided analysis for Figure~\ref{fig3} and
represent the correlations with $\langle N_{\text{part}}\rangle$
that were previously observed. In the Fokker-Planck framework,
the rise of \(b = \sqrt{B_0/\beta}\) shows that the constant
part of the diffusion coefficient develops more quickly than
its quadratic component, and the increase of \(T = B_{0}/A_{0}\)
suggests a relative strengthening of momentum-space diffusion
over drag. In high-multiplicity collisions, the hadron spectra
become more non-extensive due to the simultaneous increase
of \(c = 1 + \alpha/(2\beta)\) and \(d\), which controls the
sharpness of the transition from an exponential to a power-law
tail.

When combined, the trends in Figure~\ref{fig4} show that
the generalized Fokker–Planck solution is capable of explaining
how the system changes from small, peripheral interactions to
large, central ones, with \( \la dN_{\rm ch}/d\eta \ra \) serving
as a significant indicator for the system size and energy density.

\section{Conclusions}
\label{sec:conclu}

Using data from the PHENIX Collaboration, we investigated the
transverse momentum spectra of neutral pions from four asymmetric
small-system collisions at $\sqrt{s_{NN}}=200~{\rm GeV}$: $p$-Al,
$p$-Au, $d$-Au, and $^3$He-Au. We employed a generalized
Fokker-Planck solution (\ref{eq5}) to fit these spectra, which effectively
connects the power-law tail at high-$p_{\rm T}$ (where hard scatterings
dominate) with the exponential low-$p_{\rm T}$ part (where thermal
physics dominates). It is interesting that this single functional
form is applicable to all four collision systems and multiple centrality
bins. This indicates that the fundamental transport picture is
quite generic.

The effective temperature $T$, the energy scale $b$ that marks
the shift of the spectrum from exponential to power-law, and the
two exponents $c$ and $d$ that control the rate at which the
spectrum diminishes at high $p_{\rm T}$ were all extracted from the fits.
The effective temperature increases from $T \simeq 0.33~{\rm GeV}$
in peripheral $p$-Al to $T \simeq 0.45~{\rm GeV}$ in central $^3$He-Au
in a systematic manner as collisions become more central and
the projectile-target system gets larger. The upward trend is
consistent with both the multiplicity of charged particles and
the average number of participating nucleons. In other words:
larger and more central collisions generate
a denser, more strongly interacting medium that remains
hotter and freezes out later. According to the Fokker-Planck model, a
rise in $T$ indicates that, as the system expands, and momentum-space
diffusion becomes stronger in comparison to drag (or drift). 

The other parameters provide insight into the transport
dynamics. The crossover from thermal to hard-scattering behavior
moves to higher transverse momenta in larger collisions because
the transition scale $b$ likewise increases with centrality and
system size. One intriguing finding is that in $^3$He-Au
collisions the parameter $b$ remains nearly independent
of the centralities. This indicates that
adding more participants to the projectile a has minor
effect on the diffusion scale, which saturates after the
gold target dominates the collision geometry.

The observed increase of $b$ with an increase in
collision centrality and system size reflects an enhanced transverse
flow velocity and associated larger pressure gradients in the
expanding system in central --- compared to peripheral --- collisions.
Indeed, an increased radial flow boosts particles to higher transverse
momenta, which ``flattens'' the spectrum, shifting the transition
between an exponential and a power-law spectrum to
higher transverse momenta. Hence, a larger $b$-value is required
to reproduce this flatter slope for more central collisions in
the low-to-intermediate momentum region.

A systematic but
saturating shift in the transport coefficients is observed for
central collisions by the exponent $c$, which corresponds to
the ratio of drift and quadratic diffusion coefficients,
increasing from $p$-Al to $d$-Au and then remaining roughly
constant for $^3$He-Au. A distinct ordering is demonstrated
by the other exponent $d$, which determines how quickly the
spectrum switches from an exponential to a power-law behavior.
It is largest for $p$-Au and smallest for $d$-Au. Therefore,
the precise combination of the projectile and target, rather
than just the number of nucleons involved, determines how sharp
that transition is. 

All aspects considered, our analysis demonstrates that the
generalized Fokker-Planck solution is an efficient and cohesive
tool for investigating the strongly interacting medium that
forms in small-system asymmetric collisions, regardless of
whether that medium is a fully developed quark-gluon plasma
or something else with a comparable strong collective behavior.
Our extracted parameters provide a quantitative measure
of the non-extensive nature of the medium as well as transport
features such as drag and diffusion.

In order to figure out how general this description is,
and for the sake of a better understanding
of the phase diagram of strongly interacting matter, it is
a logical next step to apply this method to other particle
species, different collision energies, and larger systems
like nucleus-nucleus collisions.\\

\noindent
{\bf Acknowledgments:}
The authors extend their appreciation to the Deanship of Scientific
Research at Northern Border University, Arar, KSA for funding this
research work through the project number NBU-FFR-2026-2099-05.
This work was also supported by the Princess Nourah bint Abdulrahman
University Researchers Supporting Project number (PNURSP2026R443),
Princess Nourah bint Abdulrahman University, Riyadh, Saudi Arabia,
by the Ajman University Internal Research
Grant No.\ [DRGS Ref. 2025-IRG-HBS-13], and by the Mexican
funding agency UNAM-DGAPA through project PAPIIT IG100826.
\\
\end{multicols}

{\small
\begin{thebibliography}{99}
\setlength{\itemsep}{-1pt}

\bibitem{asymfree} D.J.~Gross and F.~Wilczek, 
\href{https://doi.org/10.1103/PhysRevLett.30.1343}
{Phys. Rev. Lett. \textbf{30} (1973) 1343-1346}.
H.D.~Politzer,
\href{https://doi.org/10.1103/PhysRevLett.30.1346}
{Phys. Rev. Lett. \textbf{30} (1973) 1346-1349}.

\bibitem{FGML} H.~Fritzsch, M.~Gell-Mann and H.~Leutwyler, 
\href{https://doi.org/10.1016/0370-2693(73)90625-4}
{Phys. Lett. \textbf{47} B (1973) 365-368}.

\bibitem{QCD50} F.~Gross {\it et al.}, 
\href{https://doi.org/10.1140/epjc/s10052-023-11949-2}
{Eur.\ Phys.\ J.\ C \textbf{83} 
(2023) 1125}.

\bibitem{Schmidt:2017bjt} 
C.~Schmidt and S.~Sharma,
\href{https://doi.org/10.1088/1361-6471/aa824a}
{J. Phys. G \textbf{44} 
(2017) 104002}.

\bibitem{Haegler} Ph.~H\"{a}gler, 
\href{https://doi.org/10.1016/j.physrep.2009.12.008}
{Phys. Rep. \textbf{490} (2010) 49-175}.

\bibitem{PDB} S.~Navas {\it et al.} (Particle Data Group), 
\href{https://doi.org/10.1103/PhysRevD.110.030001}
{Phys. Rev. D \textbf{110} (2024) 030001}.

\bibitem{FLAG} Y.\ Aoki {\it et al.} (Flavour Lattice Averaging
Group (FLAG)), 
\href{https://doi.org/10.1103/nfzp-p5dn}
{Phys. Rev. D \textbf{113} (2026) 014508}.

\bibitem{STAR:2019bjj} 
J.~Adam \textit{et al.} (STAR Collaboration),
\href{https://doi.org/10.1103/PhysRevC.102.034909}
{Phys. Rev. C \textbf{102} 
(2020) 034909}.

\bibitem{STAR:2003fka} 
J.~Adams \textit{et al.} (STAR Collaboration),
\href{https://doi.org/10.1103/PhysRevLett.91.172302}
{Phys. Rev. Lett. \textbf{91} (2003) 172302}.

\bibitem{ALICE:2019nrq} 
S.~Acharya \textit{et al.} (ALICE Collaboration),
\href{https://doi.org/10.1016/j.physletb.2020.135434}
{Phys. Lett. B \textbf{805} (2020) 135434}.

\bibitem{ALICE:2012ovd} 
B.~Abelev \textit{et al.} (ALICE Collaboration),
\href{https://doi.org/10.1103/PhysRevLett.109.252301}
{Phys. Rev. Lett. \textbf{109} (2012) 252301}.

\bibitem{Abbott1} 
B.P.~Abbott \textit{et al.} (LIGO Scientific, Virgo), 
\href{https://doi.org/10.1103/PhysRevLett.119.161101}
{Phys. Rev. Lett. \textbf{119} (2017) 161101}.

\bibitem{Abbott2} 
B.P.~Abbott \textit{et al.} (LIGO Scientific, Virgo, Fermi GBM, INTEGRAL,
IceCube, AstroSat Cadmium Zinc Telluride Imager Team, IPN,
Insight-Hxmt, ANTARES, Swift, AGILE Team, 1M2H Team,
Dark Energy Camera GW-EM, DES, DLT40, GRAWITA, Fermi-LAT, ATCA,
ASKAP, Las Cumbres Observatory Group, OzGrav, DWF (Deeper
Wider Faster Program), AST3, CAASTRO, VINROUGE, MASTER,
J-GEM, GROWTH, JAGWAR, CaltechNRAO, TTU-NRAO, NuSTAR,
Pan-STARRS, MAXI Team, TZAC Consortium, KU, Nordic Optical
Telescope, ePESSTO, GROND, Texas Tech University, SALT Group,
TOROS, BOOTES, MWA, CALET, IKI-GW Follow-up, H.E.S.S.,
LOFAR, LWA, HAWC, Pierre Auger, ALMA, Euro VLBI Team, Pi of Sky,
Chandra Team at McGill University, DFN, ATLAS Telescopes,
High Time Resolution Universe Survey, RIMAS, RATIR,
SKA South Africa/MeerKAT),
\href{https://doi.org/10.3847/2041-8213/aa91c9}
{Astrophys. J. Lett. \textbf{848} (2017) L12}.

\bibitem{Laermann:2003cv} 
E.~Laermann and O.~Philipsen,
\href{https://doi.org/10.1146/annurev.nucl.53.041002.110609}
{Ann. Rev. Nucl. Part. Sci. \textbf{53} (2003) 163-198}.

\bibitem{PHOBOS} 
B.B.~Back \textit{et al.} (PHOBOS Collaboration),
\href{https://doi.org/10.1016/j.nuclphysa.2005.03.084}
{Nucl. Phys. A \textbf{757} (2005) 28-101}.

\bibitem{PHENIX} 
K.~Adcox \textit{et al.} (PHENIX Collaboration),
\href{https://doi.org/10.1016/j.nuclphysa.2005.03.086}
{Nucl. Phys. A \textbf{757} (2005) 184-283}.

\bibitem{STAR} 
J.~Adams \textit{et al.} (STAR Collaboration),
\href{https://doi.org/10.1016/j.nuclphysa.2005.03.085}
{Nucl. Phys. A \textbf{757} (2005) 102-183}.

\bibitem{BRAHMS} 
I.~Arsene \textit{et al.} (BRAHMS Collaboration),
\href{https://doi.org/10.1016/j.nuclphysa.2005.02.130}
{Nucl. Phys. A \textbf{757} (2005) 1-27}. 

\bibitem{Adamczyk16} 
L.~Adamczyk \textit{et al.} (STAR Collaboration),
\href{https://doi.org/10.1103/PhysRevLett.116.062301}
{Phys. Rev. Lett. \textbf{116} (2016) 062301}.

\bibitem{STAR:2015vmv} 
L.~Adamczyk \textit{et al.} (STAR Collaboration),
\href{https://doi.org/10.1103/PhysRevLett.116.132301}
{Phys. Rev. Lett. {\bf 116} (2016) 132301}.

\bibitem{STAR:2016uxt} 
L.~Adamczyk \textit{et al.} (STAR Collaboration),
\href{https://doi.org/10.1103/PhysRevC.93.064904}
{Phys. Rev. C {\bf 93} (2016) 064904}.

\bibitem{STAR:2017akg} 
L.~Adamczyk \textit{et al.} (STAR Collaboration),
\href{https://doi.org/10.1103/PhysRevD.97.032004}
{Phys. Rev. D {\bf 97} (2018) 032004}.

\bibitem{Tawfik:2021xfc} 
A.N.~Tawfik,
16th Marcel Grossmann Meeting on Recent Developments in
Theoretical and Experimental General Relativity, Astrophysics and
Relativistic Field Theories, 5-9 July 2021, MG16, 4277-4289
(2021), doi: 10.1142/9789811269776$\_$0359.

\bibitem{qq} 
X.~Luo and N.~Xu,
\href{https://doi.org/10.1007/s41365-017-0257-0}
{Nucl. Sci. Tech. \textbf{28} 
(2017) 112}.

\bibitem{qqq} 
N.~Xu (STAR Collaboration),
\href{https://doi.org/10.1016/j.nuclphysa.2014.10.022}
{Nucl. Phys. A \textbf{931} (2014) 1-12}.
 
\bibitem{q1} 
M.~Badshah \textit{et al.},
\href{https://doi.org/10.1088/1361-6471/ad41f2}
{J. Phys. G: Nucl. Part. Phys. \textbf{51} (2024) 065109}.

\bibitem{q2} 
M.~Ajaz \textit{et al.},
\href{https://doi.org/10.1088/1674-1137/ad2a4c}
{Chin. J. Phys. \textbf{48} (2024) 053108}.

\bibitem{q3} 
H.I.~Alrebdi, M.~Ajaz, M.~Waqas, M.A.~Ahmad,  Maryam, A.M.~Quraishi,
J.H.~Baker, S.~Jagnandan and A.~Jagnandan,
\href{https://doi.org/10.1016/j.cjph.2024.02.034}
{Chin. J. Phys. \textbf{89} (2024) 1669-1677}.

\bibitem{Waqas:2022fnl} 
M.~Waqas, G.X.~Peng, M.~Ajaz, A.A.~Ismail Haj, Z.~Wazir and L.~L.~Li,
\href{https://doi.org/10.1088/1361-6471/ac6a00}
{J. Phys. G \textbf{49} (2022) 095102}.

\bibitem{Badshah:2024pyt} 
M.~Badshah, H.I.~Alrebdi, M.~Waqas, M.~Ajaz and M.B.~Ammar,
\href{https://doi.org/10.1140/epja/s10050-024-01357-9}
{Eur. Phys. J. A \textbf{60} (2024) 139}.

\bibitem{Svetitsky88} 
B.~Svetitsky, 
\href{https://doi.org/10.1103/PhysRevD.37.2484}
{Phys. Rev. D \textbf{37} (1988) 2484-2491}.

\bibitem{Banerjee88} 
A.~Banerjee and V.M.~Yakovenko,
\href{https://doi.org/10.1088/1367-2630/12/7/075032}
{New J. Phys. \textbf{12} (2010) 075032}.

\bibitem{HZheng} 
H.~Zheng and L.~Zhu,
\href{http://dx.doi.org/10.1155/2015/180491}
{Adv. in High Energy Phys. \textbf{2015} (2015) 180491}.

\bibitem{XuejiaoYin} 
X.~Yin, L.~Zhu and H.~Zheng,
\href{http://doi.org/10.1155/2017/6708581}
{Adv. in High Energy Phys. (2017) 6708581}.

\bibitem{ZZZB20} H.\ Zheng, X.\ Zhu, L.\ Zhu and A.\ Bonasera, 
\href{http://doi.org/10.1142/S0217732320501771}
{Mod. Phys. Lett. A \textbf{35} (2020) 2050177}.

\bibitem{NJP} 
A.~Banerjee and V.M.~Yakovenko, 
\href{https://doi.org/10.1088/1367-2630/12/7/075032}
{New J. Phys. \textbf{12} (2010) 075032}. 

\bibitem{Yin:2017pzp} 
X.~Yin, L.~Zhu and H.~Zheng,
\href{https://doi.org/10.1155/2017/6708581}
{Adv. High Energy Phys. \textbf{2017} (2017) 6708581}.

\bibitem{PHENIX:2021dod} 
U.A.~Acharya \textit{et al.} (PHENIX Collaboration),
\href{https://doi.org/10.1103/PhysRevC.105.064902}
{Phys. Rev. C \textbf{105} (2022) 064902}.

\bibitem{Waqas:2022nae} 
M.~Waqas, G.X.~Peng, M.~Ajaz, A.~Haj Ismail and E.A.~Dawi,
\href{https://doi.org/10.1103/PhysRevD.106.075009}
{Phys. Rev. D \textbf{106} (2022) 075009}.

\bibitem{Lao:2021wub} 
H.L.~Lao, F.H.~Liu and B.~Q.~Ma,
\href{https://doi.org/10.3390/e23070803}
{Entropy \textbf{23} (2021) 803}.

\bibitem{Waqas:2021rmb} 
M.~Waqas, G.X.~Peng and F.H.~Liu,
\href{https://doi.org/10.1088/1361-6471/abdd8d}
{J. Phys. G \textbf{48} (2021) 075108}.


\bibliographystyle{plain}
\end{thebibliography}}
\end{document}